\documentclass[twoside]{IEEEtran}

\newcommand\IEEEheadjournal[3]{

\title{Hybrid Memristor-CMOS (MeMOS) based Logic Gates and Adder Circuits}

\author{Tejinder Singh,~\IEEEmembership{Member,~IEEE}
\thanks{Manuscript received Month XX, 201X; revised Month XX, 201X; accepted Month XX, 201X. Date of publication Month XX, 201X.}
\thanks{The authors are with the Discipline of Electronics and Electrical Engineering, lovely Professional University, Phagwara 144 402, PB, India (e-mail: tejinder.singh@ieee.org)}
\thanks{Color versions of one or more of the figures in this brief are available online at http://ieeexplore.ieee.org.}
\thanks{Digital Object Identifier 10.XXXX/XXXX.201X.XXXXXXX}}

\markboth{{#3},~Vol.~X,~No.~X, Month~Year}%
{T. Singh\MakeLowercase{\textit{}}: {#1}}

\IEEEpubid{1234--5678/01 \$31.00~\copyright~2015 Publisher}

\maketitle

\IEEEpubidadjcol

}

\usepackage{cite, siunitx}
\usepackage{array}
\usepackage{booktabs}
\usepackage{upgreek}
\usepackage{multirow}
\usepackage{hhline}
\usepackage{footnote}
\usepackage[cmex10]{amsmath}
\ifCLASSINFOpdf
\usepackage[pdftex]{graphicx}
\usepackage{epstopdf}
   \else
   \fi

\makeatletter
\newcommand{\thickhline}{%
    \noalign {\ifnum 0=`}\fi \hrule height 1pt
    \futurelet \reserved@a \@xhline
}

\newcommand{\dd}{\mathop{}\!\mathrm{d}}

\newcolumntype{"}{@{\hskip\tabcolsep\vrule width 1pt\hskip\tabcolsep}}
\makeatother

\newcolumntype{M}{>{\centering\arraybackslash}}

\newcommand{\NOT}{\texttt{NOT}~}
\newcommand{\AND}{\texttt{AND}~}
\newcommand{\OR}{\texttt{OR}~}
\newcommand{\NAND}{\texttt{NAND}~}
\newcommand{\NOR}{\texttt{NOR}~}
\newcommand{\XOR}{\texttt{XOR}~}
\newcommand{\XNOR}{\texttt{XNOR}~}

\newcommand{\VCC}{\texttt{V$_{\texttt{CC}}$}~}

\renewcommand{\tt}{\texttt}

\DeclareSIUnit\lb{lb}
\DeclareSIUnit\um{\mu m}
\DeclareSIUnit\deg{\SIUnitSymbolDegree}
\DeclareSIUnit\ghz{GHz}
\DeclareSIUnit\khz{KHz}
\DeclareSIUnit\mhz{MHz}
\DeclareSIUnit\db{dB}
\DeclareSIUnit\mpa{MPa}
\DeclareSIUnit\gpa{GPa}
\DeclareSIUnit\fF{fF}
\DeclareSIUnit\pF{pF}
\DeclareSIUnit\pH{pH}
\DeclareSIUnit\ng{ng}

\begin{document}
\IEEEheadjournal{Hybrid Memristor-CMOS (MeMOS) based Logic Gates and Adder Circuits}{}{Journal Name}

\begin{abstract}
Practical memristor came into picture just few years back and instantly became the topic of interest for researchers and scientists. Memristor is the fourth basic two--terminal passive circuit element apart from well known resistor, capacitor and inductor. Recently, memristor based architectures has been proposed by many researchers. In this paper, we have designed a hybrid Memristor-CMOS (MeMOS) logic based adder circuit that can be used in numerous logic computational architectures. We have also analysed the transient response of logic gates designed using MeMOS logic circuits. MeMOS use CMOS 180\:nm process with memristor to compute boolean logic operations. Various parameters including speed, ares, delay and power dissipation are computed and compared with standard CMOS 180\:nm logic design. The proposed logic shows better area utilisation and excellent results from existing CMOS logic circuits at standard 1.8\:V operating voltage.
\end{abstract}


\section{Introduction}
\begin{figure}[!t]
\centering
\includegraphics[width=3.2in]{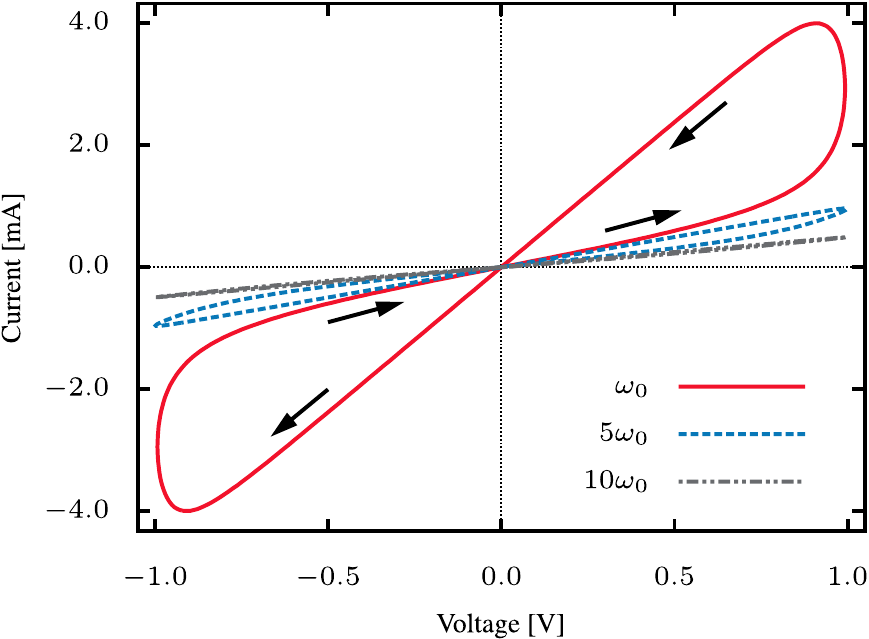}
\caption{Pinched hysteresis loop of memristor represents the Current--Voltage characteristics of a linear ion drift TiO$_2$ memristor. $\omega_0$ is the frequency of input signal under the applied voltage bias of v$_0\sin$($\omega_0$t). The respective curves for $5\omega_0$ and $10\omega_0$ is shown.}
\label{ivchar}
\end{figure}

\IEEEPARstart{M}{emristor} is the well known device now-a-days, that captured the interest of researchers and scientists when HP Labs realized a practical physical device in 2008. It all started back when L. Chua in 1971 on the basis of symmetry forefront, postulated memristor\footnote{The term Memristor and Memristive Systems can be used interchangeably.} (short of memory resistor) as the fourth basic fundamental circuit element.~\cite{balke2010real,biolek2012analytical,biolek2012computation,blanc1971electrocoloration,borghetti2009hybrid,borghetti2010memristive} Memristor basically connects electric charge~$q$ and magnetic flux~$\phi$ as voltage~$V$ and current~$I$ is connected by resistor, magnetic flux~$phi$ and current~$I$ by inductor and voltage~$V$ and charge~$q$ is connected by capacitor. The relation charge~$q$ and magnetic flux~$\phi$ was missing as per Chua stated. Chua demonstrated that memristors can be characterised by pinched hysteresis loop as shown in Fig.~\ref{ivchar}. In theory the memristor term was extended to memristive devices in 1976 by S. Kang. Characterisation of memristors require two equations instead of one.~\cite{chan1987impact,chang2011short,chang2011synaptic,chanthbouala2012ferroelectric,cheng2011nanoelectronics,chua1959memristors}

In 2008, HP Labs realised memristor that consists of TiO$_2$ thin film sandwiched between two platinum electrodes on both sides. The response of the linear ion drift memristor is shown in Fig.~\ref{ivchar} for frequency $\omega_0$, $5\omega_0$ and $10\omega_0$. Now, after few years of research, memristors are considered as one of the best alternative to current generation CMOS technology. Memristors are basically the devices with varying resistance that depends on the previous state of the device. Memristors can be voltage or current driven. Memristors can be used for memory implementation, where the logic bits are stored as resistance states. Various applications has been proposed by researchers recently that includes neuromorphic applications and use in analog circuits. 

One major area of interest is the logic computation by using memristors. Researchers has proposed different methods of logic computation. One of the primary methodology that is most regarded is the material implication using memristor. Some has proposed integration with CMOS logic to compute various logical operations. Material implication logic shows promising results but need more computational steps in performing logic. The major constrain with material implication is the designing of read/write circuits as the logic is completely different from boolean logic. Moreover, it is not compatible with current generation CMOS technology.~\cite{corinto2012boundary,deng2013rram,devolder2008single} 

\IEEEpubidadjcol

Hybrid Memristor-CMOS logic is a hybrid of both memristors and CMOS. Using this implementation scheme, we can compute logic and the outputs can also be represented by voltage levels. We coined the term `MeMOS' that better suits Memristor-CMOS hybrid integration. In this approach, \AND and \OR logic can be computed using the memristors only and CMOS inverter is used to get \NOT operation as the operation \NOT is not possible with memristors only.

Many mathematical models of memristor presented by researchers. We have chosen TeAM (Threshold Adaptive Memristor Model) for our study as, the model has current threshold parameters and provides realistic modeling for logic implementation. Although, there are many proposed models for logic computation but in our study we have kept the voltage level at standard 1.8\:V and designed logic gates and thus full adder. We have also simulated the response of logic gates and full adder with CMOS only by keeping the same MOS parameters. This study will give a fair enough idea that by using MeMOS logic, we can save much more area than the current CMOS implementation require. There are different approaches to design adder but we have chosen the most basic one for the comparison sake.\cite{driscoll2011chaotic,elwakil2013simple,fagg2007biomimetic,fouda2012effect,fouda2013generalized} Different advantages and issues with this logic is also discussed in next sections.

The paper is organised as follows: Section II describes the modeling of TeAM memristor model. The schemeatic of \AND and \OR logic computation models are described in Section III. Section IV describes the logic gates designing and transient response analysis. Full adder circuit is described in Section V and the comparison of parameters like delay, rise time and fall time is given in Section VI followed by Section VII which summarise the paper and future work is given in the same section.   

\section{Memristor Modelling}
The memristor was postulated by L. Chua in 1971 based on the symmetry forefront. The operation of memristor depends on the history or the previous state of the device. The memristor is considered as varying resistor with varying resistance as memristance $M$ of device. The memristance $M$ of the memristor depends on the total current passed through the device. When the voltage or current is removed from the source, the memristor retains its state and by applying lower value of voltage or current the previous state can be read easily.\cite{ho2009nonvolatile,ho2011dynamical,hu2011self,iu2011controlling} In 1976 L. Chua and S. Kang generalised the concept of memristor to a broader class of nonlinear dynamical systems.\cite{kavehei2012analytical} They called these systems as the memristive systems, the current-controlled, time-invariant memristor can be described by the equations as \cite{kim2012neural} 

\begin{equation}
v = \mathcal{R} \left( w, i \right)i
\end{equation}

\begin{equation}
\frac{\dd w}{\dd t} = f \left( w, i \right)
\end{equation}

\noindent
where $w$ is a set of state variables of the device and $f$ and $\mathcal{R}$ are the explicit functions of time. $v$ and $i$ are voltage and current with respect to time respectively.

\begin{figure}[b!]
\centering
\includegraphics[width=2in]{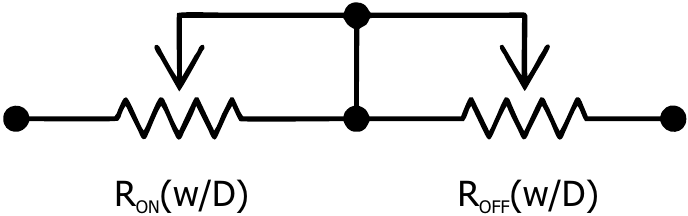}
\caption{Schematic of linear ion drift memristor as proposed by HP Labs}
\label{memristor_model}
\end{figure}

The most popular and common memristor model is the Linear ion drift model that is based on the memristor characteristics described by HP Labs in 2008. In linear ion drift model, a device of physical width $D$ is considered that has two regions viz. dopes and undoped as shown in Fig.~\ref{memristor_model}. A region of width $w$ has high dopants concentration. $w$ also acts as the state variable of the device. The dopants are oxygen vacancies that are TiO$_{2-x}$ for the case of TiO$_2$ based memristor. The other region of width $D-w$ is generally an oxide region with the dopant with higher conductance than that of oxide region with mobility of ions is described by $\mu_v$. \cite{laiho2010cellular,lam1972formulation} The device is modelled as two series connected resistors. For the assumption that linear ion drift is in uniform field and ions have similar average ion mobility $\mu_v$, the equations~(1) and (2) can be represented by \cite{long2011understanding}

\begin{equation}
v \left( t \right) = \left( \mathcal{R}_{\textrm{ON}} \frac{w(t)}{D} + \mathcal{R}_{\textrm{OFF}} \left( 1 - \frac{w(t)}{D} \right) \right) i \left( t \right)
\end{equation}

\begin{equation}
\frac{\dd w}{\dd t} = \mu_v \frac{\mathcal{R}_{\textrm{ON}}}{D} i \left( t \right) 
\end{equation} 

Equations~(3) and (4) yields the following equation for state variable $w(t)$ as

\begin{equation}
w \left( t \right) = \mu_v \frac{\mathcal{R}_{\textrm{ON}}}{D} q \left( t \right) 
\end{equation} 

Plugging the values from equation~(5) into equation~(3), we can compute the memristance of the system, for the condition $\mathcal{R}_{\textrm{ON}} \ll \mathcal{R}_{\textrm{OFF}}$ that further reduces to the equation
 
\begin{equation}
M \left( q \right) = \mathcal{R}_{\textrm{OFF}} \left( 1 - \frac{\mu_v \mathcal{R}_{\texttt{ON}}}{D^2} q \left( t \right) \right) 
\end{equation} 

\noindent
where, $M(q)$ is the memristance of the memristive system.

\begin{figure}
\centering
\includegraphics[width=3.2in]{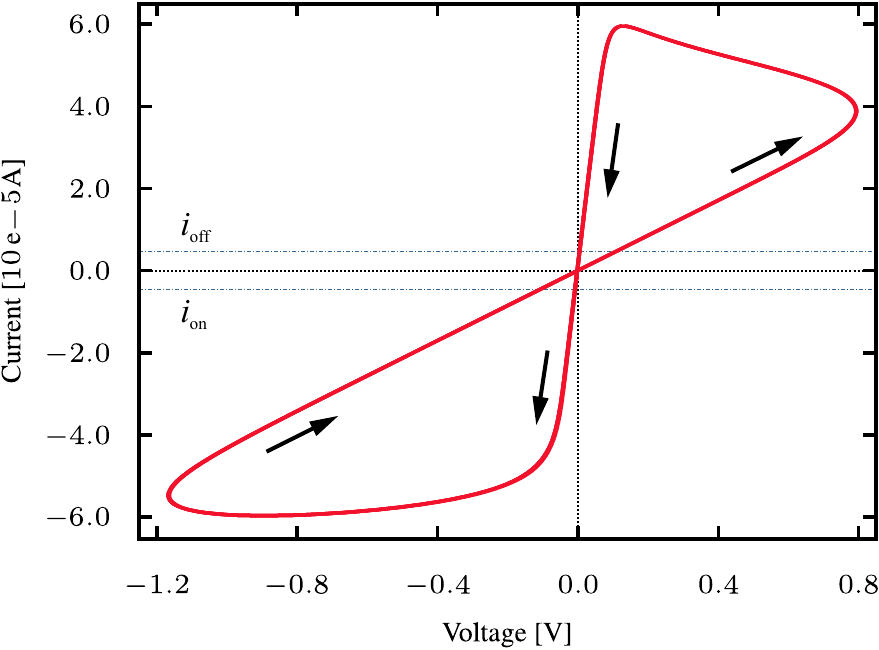}
\caption{Current--voltage characteristics simulated for TeAM model and Kvatinsky window using MATLAB memristor tool. The parameters used to plot are: $
\alpha_{\textrm{on}}~\textrm{and}~\alpha_{\textrm{off}} = 5,
i_{\textrm{on}}=-5.0\:\mu \textrm{A}, 
k_{\textrm{on}}=-0.001, 
a_{\textrm{on}}=2.3\:\textrm{nm}, 
x_{\textrm{on}}=3\:\textrm{nm},
\mathcal{R}_{\textrm{ON}}=100:\Omega,
\mathcal{R}_{\textrm{OFF}}=100\:\textrm{K}\Omega,
i_{\textrm{off}}=5.0\:\mu \textrm{A},
k_{\textrm{off}}=0.001, 
a_{\textrm{off}}=1.2\:\textrm{nm},
x_{\textrm{off}}=0,
\textrm{and window function}~f_{\textrm{on}}~\textrm{and}~f_{\textrm{off}}~\textrm{are as given in eq~(15).} 
$}
\label{team_char}
\end{figure}

To describe the physical behavior of the device, various memristor models have been proposed. The models proposed are mostly deterministic and mostly do not consider the stochastic switching behavior. The threshold adaptive memristor model as described in (7) demonstrates that the memristors have a current threshold and have an adaptive nonlinearity. For this model, the equation~(2) becomes

\begin{equation}
\frac{\dd x \left( t \right)}{\dd t} = \left\{ \,
    \begin{IEEEeqnarraybox}[][c]{l?s}
      \IEEEstrut
	      k_{\textrm{off}} \left( \frac{i(t)}{i_{\textrm{off}}} - 1 \right)^{\alpha_{\textrm{off}}} \cdot f_{\textrm{off}} \left( x \right) & \textit{for} $0\:\textless\:i_{\textrm{off}}\:\textless\:i$, \\
      0 & \textit{for} $i_{\textrm{on}}\:\textless\:i\:\textless\:i_{\textrm{off}}$, \\
      k_{\textrm{on}} \left( \frac{i(t)}{i_{\textrm{on}}} - 1 \right)^{\alpha_{\textrm{on}}} \cdot f_{\textrm{on}} \left( x \right) & \textit{for} $i\:\textless\:i_{\textrm{on}}\:\textless\:0$,
      \IEEEstrut
    \end{IEEEeqnarraybox}
\right.
  \label{eq:example_left_right1}
\end{equation}
	      
\noindent
where the current threshold parameters are defined by $i_{\textrm{on}}$ and $i_{\textrm{off}}$, $\alpha_{\textrm{on}}$ and $\alpha_{\textrm{off}}$ parameters define the adaptive nonlinearity of device, $k_{\textrm{on}}$ and $k_{\textrm{off}}$ are the fitting parameters of memristor, and $f_{\textrm{on}} \left( x \right)$ and $f_{\textrm{off}} \left( x \right)$ are the window functions. The voltage of the memristor model $v$ described in equation~(1) can be defined for the threshold adaptive memristor model as per equation~(7) as  \cite{muthuswamy2009memristor,nickel2011memristor,pazienza2011teaching}

\begin{equation}
v \left( t \right) = \left( \mathcal{R}_{\textrm{ON}} + \frac{\mathcal{R}_{\textrm{OFF}} - \mathcal{R}_{\textrm{OFF}}}{x_{\textrm{off}} - x_{\textrm{on}}} \left( x - x_{\textrm{on}}\right) \right) \cdot i \left( t\right)
\end{equation}

\noindent
where $\mathcal{R}_{\textrm{ON}}$ and $\mathcal{R}_{\textrm{OFF}}$ are the minimum and maximum resistance of the device respectively, and $x_{\textrm{on}}$ and $x_{\textrm{off}}$ are the internal state variable $x$'s minimum and maximum allowed value for the memristor. The current-voltage characteristics simulated using TeAM model is shown in Fig.~\ref{team_char}. We have chosen the TeAM model for its explicit current-voltage relationship and memristance deduction and it shows matching memristive system definition as linear ion drift memristor model shows. This model is more generic and provides accuracy comparing practical memristive devices. \cite{chua1969introduction,chua1971memristor,chua1976memristive,corinto2011nonlinear,corinto2012boundary} The important aspect of TeAM model is the existence of threshold because it not only accurately characterises the Simmons tunnelling barrier model but also various different memristor models.

The linear ion drift behaviour fit by threshold adaptive memristor model can be given as

\begin{equation}
k_{\textrm{on}} = k_{\textrm{off}} = \mu_v \frac{\mathcal{R}_{\textrm{ON}}}{D}i_{\textrm{on}}
\end{equation} 

\begin{equation}
\alpha_{\textrm{on}} = \alpha_{\textrm{off}} = 1
\end{equation} 

Linear ion drift memristors does not have any current threshold hence the equation for $i_{off}$ and $i_{on}$ approaches to $0$ as \cite{greenlee2013comparison}

\begin{equation}
i_{\textrm{on}} = i_{\textrm{off}} \rightarrow 0
\end{equation} 

\begin{equation}
x = D - w
\end{equation} 

\begin{equation}
x_{\textrm{on(off)}} = D(0)
\end{equation} 

The window function of the TeAM model for the undoped region state variable $x_{\textrm{on}} \leq x \leq x_{\textrm{off}}$ can be defined as

\begin{equation}
f_{\textrm{on} \left( \textrm{off} \right)} = \exp \left\{ -\exp \left( \frac{| x - x_{\textrm{on} \left( \textrm{off} \right)} |}{w_c} \right) \right\}
\end{equation} 

Based on the given merits of TeAM model, \cite{kvatinsky2011memristor,kvatinsky2011verilog,kvatinsky2012models,kvatinsky2012mrl,kvatinsky2013team} we have chosen this model for logic design integrated with CMOS technology. From the fabrication point of view, memristors are compatible with current generation standard CMOS technologies. The memristors are relatively smaller in size ($\approx$ 3\:nm) and thus can be fabricated with the similar techniques used for processing the in-between metal cross-layer via. Memristors are basically thin oxide sandwiched in metal layers, whereas the oxide shows memristive effect between electrodes. Memristors offer higher density of logic elements per unit area and hence can be used to design much more logic functions on same chip area.

\section{Logic Using Memristors}

\begin{figure}[!t]
\centering
\includegraphics[width=3.5in]{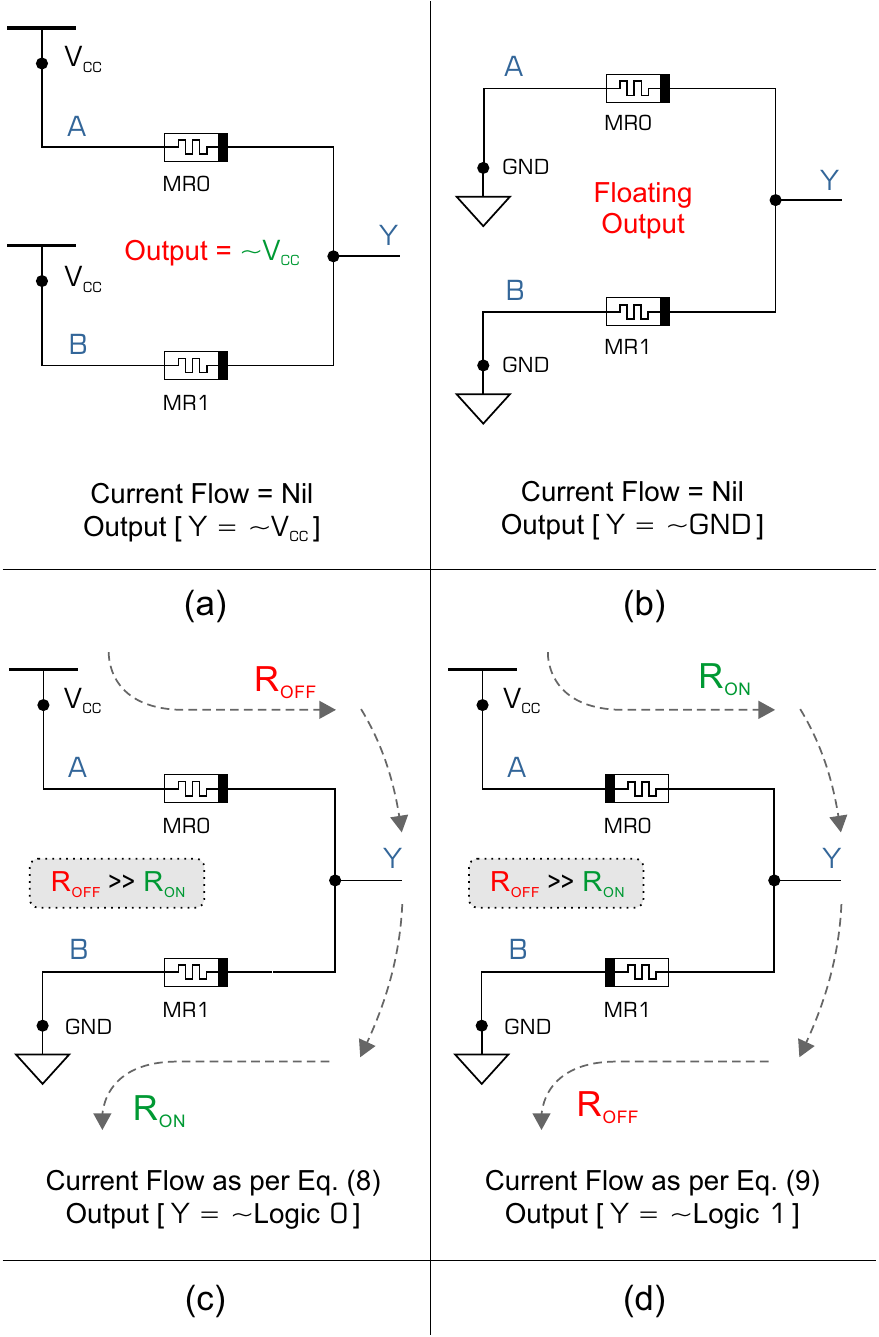}
\caption{Logic computation using Memristors. \texttt{V$_{\texttt{CC}}$} is considered as Logic \texttt{1} and \texttt{GND} as Logic \texttt{0}. \texttt{MR0} and \texttt{MR1} memristors are used to demonstrate 2--input \AND logic evaluation. (a) shows the case when both the inputs are tied to \texttt{V$_{\texttt{CC}}$} i.e., \texttt{A=1} and \texttt{B=1}, the output is \texttt{V$_{\texttt{CC}}$} = Logic \texttt{1}. (b) represents the case when any one of the input is at logic \texttt{1} and other at logic \texttt{0}. In this case, current flows from \texttt{V$_{\texttt{CC}}$} to \texttt{GND} increasing the resistance of one memristor (\texttt{MR0}) and decreasing the resistance of other memristor (\texttt{MR1}) as shown. R$_{\textrm{OFF}}$ $\gg$ R$_{\textrm{ON}}$ and thus the output \texttt{Y=0} can be computed using voltage divider rule given in Eq. (8). (c) represents the case when both the inputs are at logic \texttt{0} i.e., \texttt{A=0} and \texttt{B=0}, the output \texttt{Y=0} is floating and thus no current flows in circuit. (d) shows the truth table for \AND logic gate, where \texttt{MR0} and \texttt{MR1} is treated as inputs \texttt{A} and \texttt{B}.}
\label{currentflow}
\end{figure}

From the modeling of memristor, it is clear that the memristors exhibit varying resistance when current flows into the device or out of the device. The change in resistance $\Delta \mathcal{R}$ with respect to the direction of current flow $i \left( t \right)$ is shown in the Fig.~\ref{currentflow}. The thick black line in memristor symbol represents the polarity of the device. 

The basic boolean logic operations \AND and \OR can be analysed using memristors. Although, many researchers have reported the material implication logic using memristor but that is not compatible with current generation CMOS process. Material implication works on the state variable, the inputs and outputs are the states of memristors instead of the voltages that are required for signal propagation in CMOS process. So to integrate memristor with CMOS and to work with same voltage levels, there is a need of hybrid Memristor-CMOS (MeMOS) logic. 

In this logic, the voltages are used as logic state. Memristors can only be used as computational element rather than computational cum storage element as can be used using material implication logic. 

\begin{figure}[!b]
\centering
\includegraphics[width=3.2in]{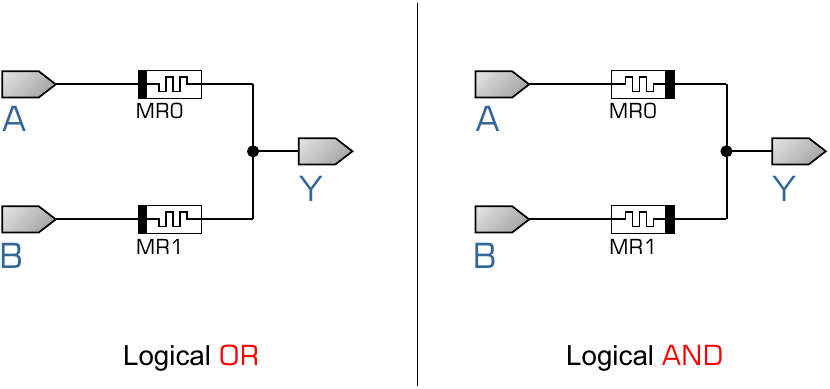}
\caption{Logical \OR and \AND operation using memristors as voltage divider}
\label{and_or}
\end{figure}

As per the current--voltage characteristics of memristor shown in Fig.~\ref{ivchar} and~\ref{team_char}, the basic idea of using memristors for logic computation is its property of varying resistance with respect to the direction of current flow through the memristor. Fig.~\ref{team_char} shows that the resistance of memristor varies depending on the direction of current flow. By using this, we can create a voltage divider circuit as reported in~\cite{borghetti2010memristive}. Fig.~\ref{and_or} shows the schematic of two input \OR and \AND gate circuit designed using memristors. We just need to change the polarity of memristors to get the correct logic value. 

Fig.~\ref{currentflow} describes the computation of \AND operation for all input cases of a two input \AND gate using memristors only. For case \texttt{A=1} and \texttt{B=1}, both the inputs are tied to \texttt{V$_{\texttt{CC}}$} i.e., at logic \texttt{1}. As described in Fig.~\ref{currentflow}(a), no current flow through the circuit and the output in this case is logic = \texttt{V$_{\texttt{CC}}$} or \texttt{1}. For the case of \texttt{A=0} and \texttt{B=0} as shown in Fig.~\ref{currentflow}(c) it can be considered that the inputs are at logic \texttt{0}, the output should also be logic \texttt{0}. Again there is no current flow through the circuit, the same logic appears at output node \texttt{Y}. These two cases are same for \OR logic also. Even with the reverse polarity of memristors, the output remains same.

For the case when any of input is at logic \texttt{1} and other at logic \texttt{0} as shown in Fig.~\ref{currentflow}(b). In this case, input \texttt{A=1} and \texttt{B=0} the current flows through \texttt{V$_{\texttt{CC}}$} to \texttt{GND}. When current passes from memristor \texttt{MR0}, the resistance of that memristor increases to $\mathcal{R}_{\textrm{OFF}}$, the resistance of memristor \texttt{MR1} decreases to $\mathcal{R}_{\textrm{ON}}$ and current leave through \texttt{GND} node. Resistance of memristors are $\mathcal{R}_{\textrm{OFF}} \gg \mathcal{R}_{\textrm{ON}}$. By this way, we get two resistors $\mathcal{R}_{\textrm{OFF}}$ and $\mathcal{R}_{\textrm{ON}}$ with different values. Thus as per the voltage divider rule, we get output \texttt{Y=0} that completes the logic for \AND gate as per truth table shown in Fig.~\ref{currentflow}(d).

The calculation of output voltage at \texttt{Y} for the voltage divider circuit can be determined as

\begin{equation}
\texttt{Y} = \texttt{V}_{\texttt{CC}} \times \frac{\mathcal{R}_{\textrm{ON}}}{\mathcal{R}_{\textrm{ON}} + \mathcal{R}_{\textrm{OFF}}}
\end{equation}

$\mathcal{R}_{\textrm{OFF}}$ is significantly higher than $\mathcal{R}_{\textrm{ON}}$ we can simplify the equation as

\begin{equation}
\texttt{Y} = \texttt{V}_{\texttt{CC}} \times \frac{\mathcal{R}_{\textrm{ON}}}{\mathcal{R}_{\textrm{OFF}}} \ll \texttt{V}_{\texttt{CC}} \approx \texttt{GND}
\end{equation}

When the polarity of the memristors \texttt{MR0} and \texttt{MR1} is reversed, the circuit behaves as \OR gate. For inputs \texttt{A=0, B=0} and \texttt{A=1, B=1} the output \texttt{Y} get the value \texttt{0} and \texttt{1} as no current flow through the circuit and the behavior remains same as in the case of \AND gate. Fig.~\ref{currentflow} (d) shows the case when any one of the input is at logic \texttt{1} and other at logic \texttt{0}. Current flows through the \texttt{V$_{\texttt{CC}}$} towards memristor \texttt{MR0}. As the memristor is in reverse polarity arrangement, the resistance of the memristor decreases to $\mathcal{R}_{\textrm{ON}}$ and thus the voltage shows up at output node \texttt{Y=V$_{\texttt{CC}}$}. The output becomes logic \texttt{1} when any of the input is at logic \texttt{1}. The resistance of other memristor \texttt{MR1} increases to $\mathcal{R}_{\textrm{ON}}$ and the condition remains same $\mathcal{R}_{\textrm{OFF}} \gg \mathcal{R}_{\textrm{ON}}$ because these are the fixed values. The output can be determined for \OR gate using the voltage divider rule as  

\begin{equation}
\texttt{Y} = \texttt{V}_{\texttt{CC}} \times \frac{\mathcal{R}_{\textrm{OFF}}}{\mathcal{R}_{\textrm{ON}} + \mathcal{R}_{\textrm{OFF}}} \approx \texttt{V}_{\texttt{CC}}
\end{equation}

\AND and \OR gates can be implemented using memristors only. By this topology, even `$n$' input gates can be implemented using memristors. But the primary issue is incomplete logic family. Without \NOT operation, it is not possible to implement boolean functions. CMOS inverter can be used to implement \NOT operation. The CMOS inverter is designed using $180$\:nm process technology. The operating voltage for the CMOS inverter is \SI{1.8}{V}. We kept the same parameter of memristors as used to simulate the current--voltage characteristics shown in Fig.~\ref{team_char}. There is an advantage of using hybrid MeMOS logic in terms of level restoration. As per the voltage divider, the output of the memristor based gate depends on the value of $\mathcal{R}_{\textrm{OFF}}$ and $\mathcal{R}_{\textrm{ON}}$. Even if there is a large difference in these two values then also the output degrades a little bit as in the denominator term of equation~(15) and (17), $\mathcal{R}_{\textrm{ON}}$ is added with $\mathcal{R}_{\textrm{OFF}}$. In the case of different inputs, we get $0.996$ \texttt{V$_{\texttt{CC}}$} at output. Hence, while cascading memristor stages, the output level decreases. PMOS is always tied to \texttt{V$_{\texttt{DD}}$}, so when signal pass through inverter, the logic level is retrieved. But in certain cases, we need \texttt{BUFFER} to restore the logic level.

Static power dissipation, delay and large area consumption are some of the primary trade-offs of integrating CMOS with memristors. Memristor layer can be fabricated on top of CMOS layer.

\begin{figure}[!b]
\centering
\includegraphics[width=3.2in]{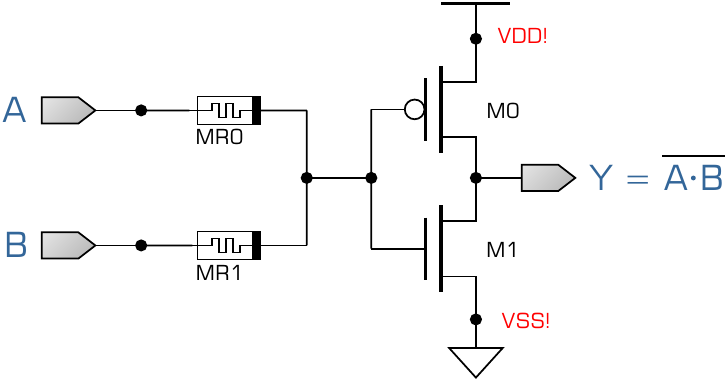}
\caption{Schematic of \NAND gate using Hybrid Memristor-CMOS logic. The memristors are in the configuration to provide \AND operation and CMOS \NOT gate is used at the output to get \NAND operation.}
\label{NAND}
\end{figure}

\begin{figure}[!b]
\centering
\includegraphics[width=3in]{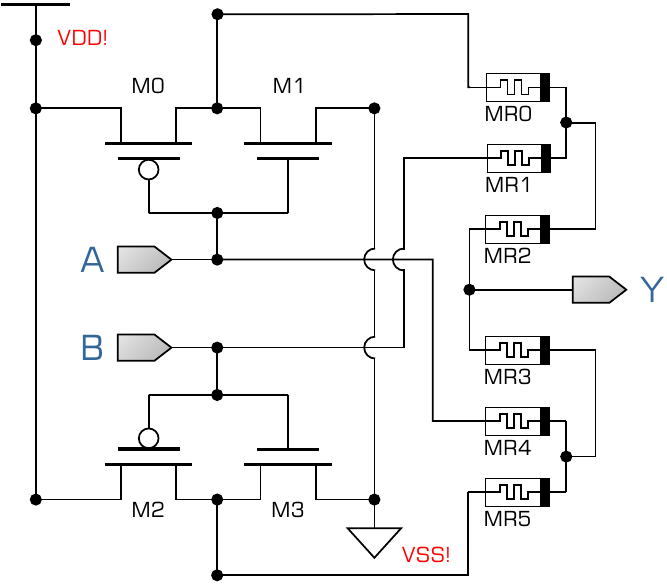}
\caption{Transient response of \XOR gate till 4 nm using Hybrid Memristor-CMOS logic shows degraded output due to the cascading of stages in \XOR implementation. The logic can be levelled by using a buffer.}
\label{XOR}
\end{figure}

\section{Logic Gates Designing using MeMOS Logic}
As described in previous section, \NOT operation is not possible with memristors. So, to get the complete logic family we should add CMOS inverter at the output of \AND gate to get \NAND operation and similarly \NOR operation can also be implemented. The operating voltage is kept at same level of \SI{1.8}{V} for all the designed gates. The schematic of \NAND or \NOR gate (reversed memristor in this case) is shown in Fig.~\ref{NAND}. Similarly \XOR gate can be designed using these approaches as shown in Fig.~\ref{XOR}

\begin{figure}[!t]
\centering
\includegraphics[width=3in]{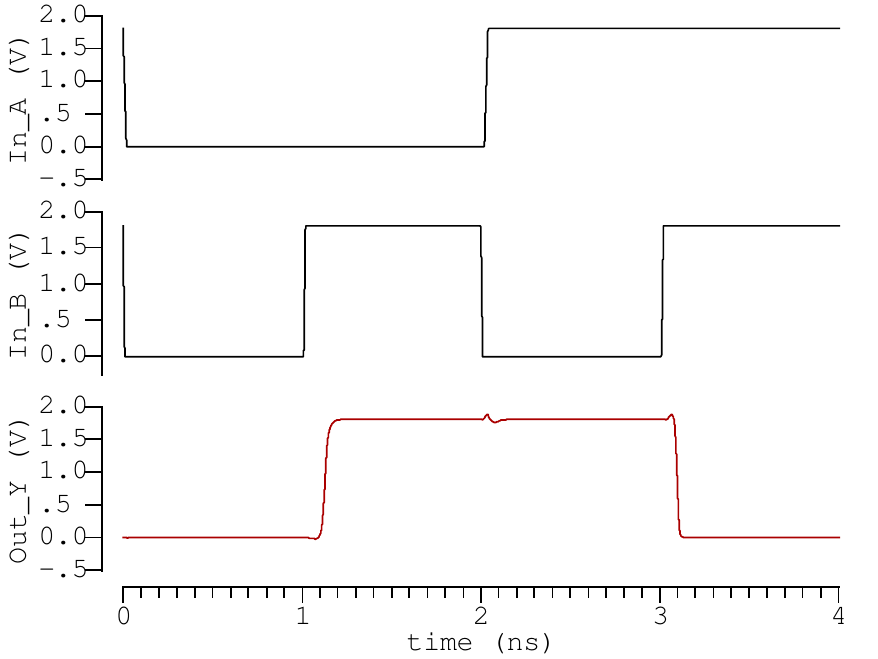}
\caption{Transient response of \XOR gate. The output of \XOR gate is levelled using a buffer showing correct logic output. The output of NAND is taken by using a \NOT gate in front of \AND gate.}
\label{XORtran}
\end{figure}

The transient response of designed \XOR gate is shown in Fig.~\ref{XORtran}. The delay $d$, rise time $t_r$ and fall time $t_f$ is kept at \SI{1.0}{ps}. Pulse signal with  width of \SI{1.0}{ns} is given at input. Cadence Virtuoso is used for designing the schematics and Spectre simulator is used to plot the transient response. The output shown for different gates is logically correct except \XOR or \XNOR gate. 

Designing of \XOR or \XNOR gate leads to logic degradation as shown in Fig.~\ref{XORtran}. The reason of degradation is explained in the previous section, it is because of the voltage divider circuit. As designing an \XOR or \XNOR gate needs cascaded stages as shown in the schematic of \XOR gate in Fig.~\ref{XOR}, in this case a \tt{BUFFER} can be used to restore the level of output voltages. The output of \XOR or \XNOR gate with \tt{BUFFER} at output is shown in Fig.~\ref{XORtran}. The level after adding \tt{BUFFER} is restored at \VCC $\approx$ \SI{1.8}{V}. Although, even in CMOS process, cascading stages need \tt{BUFFER} for level restoration but with extra area overhead.

By using MeMOS logic gates, any digital logic circuit can be implemented. We have evaluated the transient response of the these gates with the CMOS logic and found that the gates designed with MeMOS logic shows improved performance. 

\subsection{Adder Circuits}

\begin{figure}[!b]
\centering
\includegraphics[width=3.5in]{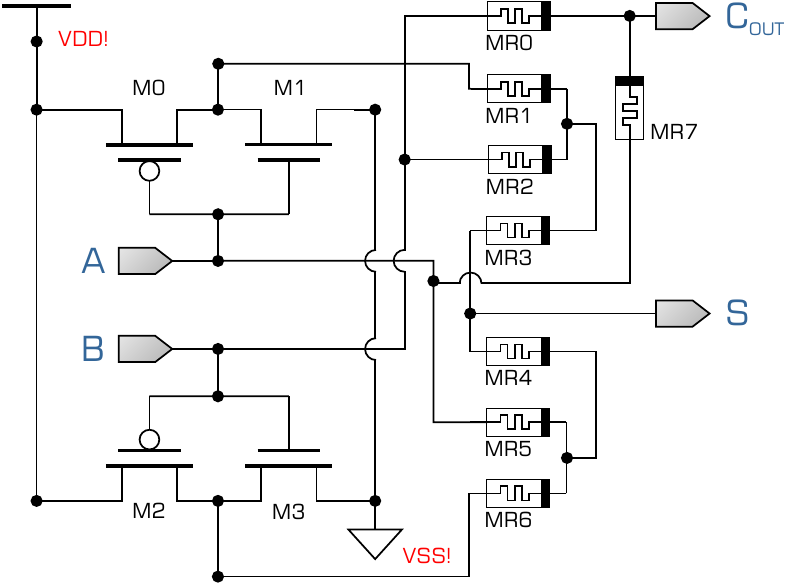}
\caption{Half Adder circuit implementation using Hybrid Memristor-CMOS logic. The circuit consumes same areas as of one \XOR gate with the inclusion of just two additional memristors.}
\label{halfadder}
\end{figure}

\begin{figure*}[!t]
\centering
\includegraphics[width=6.5in]{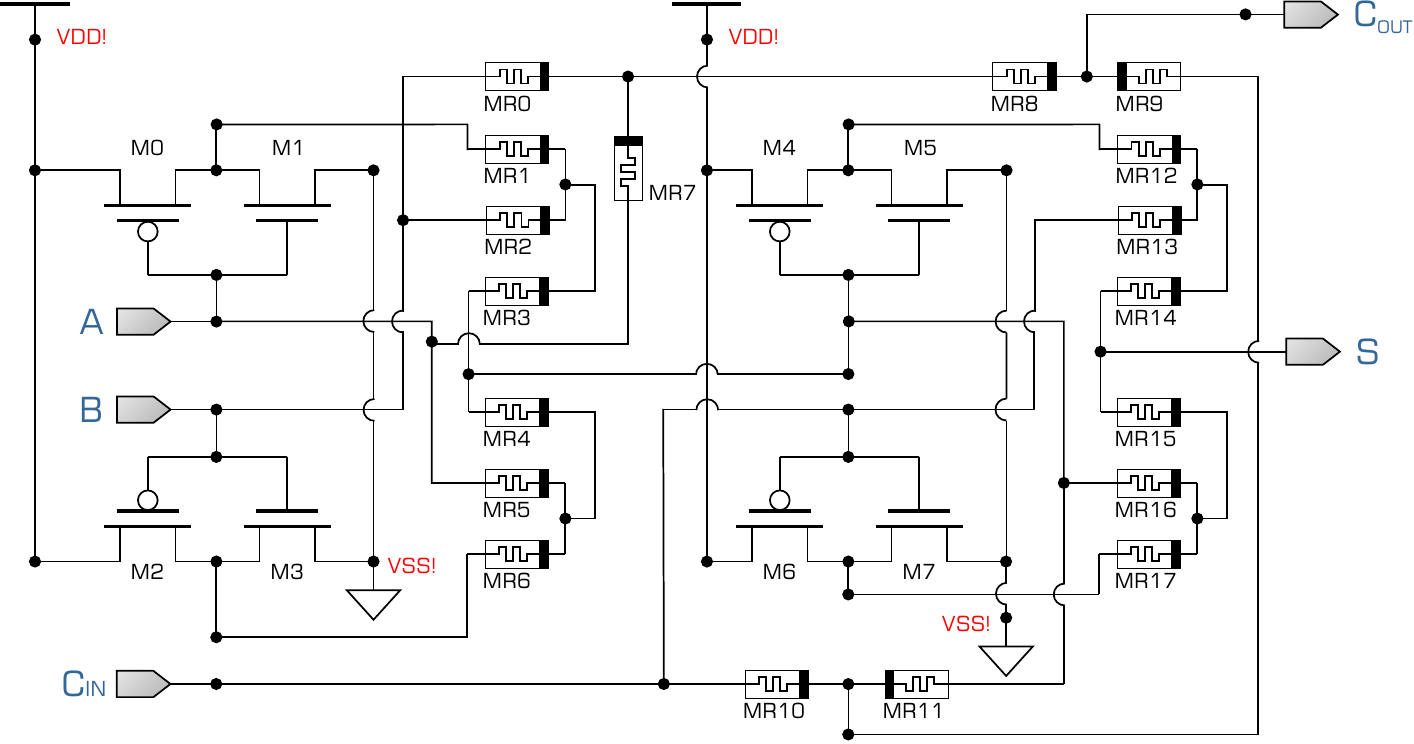}
\caption{Full Adder circuit implementation using Hybrid Memristor-CMOS logic. The circuit is implemented with 8 MOSFETS. BUFFER is required at the outputs to correct the logic degradation problem.}
\label{fulladder}
\end{figure*}

Using the gates designed using MeMOS logic, we can further extend the circuits towards the basic building block of any computation i.e., adder logic. Adders are used in different configurations to perform addition, subtraction, multiplication or division of bits. In this section, we have reported half adder, full adder and 8-bit adder circuit. Fig~\ref{halfadder} shows that by adding just two memristors in the circuit of \XOR gate, half adder can be implemented. 

In Fig.~\ref{XOR}, memristor \texttt{MR0} and \texttt{MR1} completes an \AND gate operation that is same with the memristor \texttt{MR4} and \texttt{MR5}, while memristor \texttt{MR2} and \texttt{MR3} fulfils an \OR gate. Thus connecting it in a standard fashion that leads to \XOR gate implementation. From the XOR gate, we can add one more \AND gate at the end of \XOR gate to form half adder as shown in Fig.~\ref{halfadder}. In this the memristor \texttt{MR0} and \texttt{MR7} forms \AND gate after the \XOR gate to complete the half adder operation. Similarly, A one bit full adder circuit using MeMOS logic is shown in Fig.~\ref{fulladder} that is designed using two similar half adders and memristors \tt{MR0} and \tt{MR11} act as \OR gate in the circuit to fulfil the \tt{CARRY} operation. The transient response of the adder circuit is shown in Fig.~\ref{fulladdertran}.

\subsection{Advantages over Implication Logic}
Material implication recently get hype when using memristor based \tt{IMPLY} can perform all logic tasks. \cite{hu2011self,iu2011controlling} Although, circuits designed using \tt{IMPLY} gate shows significance advantages like high speed operation, smaller area overhead and lower power consumption but except just the circuits, \tt{IMPLY} operation works on internal state resistance of memristors. External read/write circuitry is required to implement \tt{IMPLY} based logic to current generation CMOS logic a circuit is required to convert memristor's state into voltage levels for further computation. The read/write circuitry consumes much more area because of complete CMOS implementation and then due to extra circuitry, it leads to overall performance degradation of the circuit due to bottlenecks of CMOS technology.

MeMOS logic uses hybrid of both the technologies. The major advantage of using MeMOS logic is to integrate the circuits with different circuits designed by CMOS logic only. Then the input/output signals work on same voltage level that is acceptable by CMOS logic, hence the need of extra circuity diminishes. Although, it consumes more area than \tt{IMPLY} logic based circuits but considering external read/write circuitry in \tt{IMPLY} logic, MeMOS logic still has advantages over CMOS logic and close to \tt{IMPLY} logic. For Full adder circuit, two half adders are cascaded to get the desired full adder circuit. The performance is analysed for the given circuit.

The critical path \index{Critical Path}can examine all the transitions. In Fig.~\ref{halfadder}, the circuit is designed that is almost identical to \XOR gate. The circuit area is totally similar to \XOR gate designed using hybrid memristor-CMOS logic as shown in Fig.~\ref{XOR}, but the circuit act like as half-adder. Thus with full adder circuit implementation, by using just 8 MOSFETs the adder circuit works as desired. The degradation problem can be solved by adding \texttt{BUFFER} at the output.

\section{Performance Analysis}
Transient response of MeMOS based adders are computed and various parameters like rise time, fall time, delay are analysed and compared with current generation CMOS technology. Table~\ref{hybrid_parameters} shows the performance analysis of various logic gates using MeMOS logic and Table~\ref{cmos_parameters} shows the performance analysis of gates using CMOS logic. 

\begin{figure}[!t]
\centering
\includegraphics[width=3in]{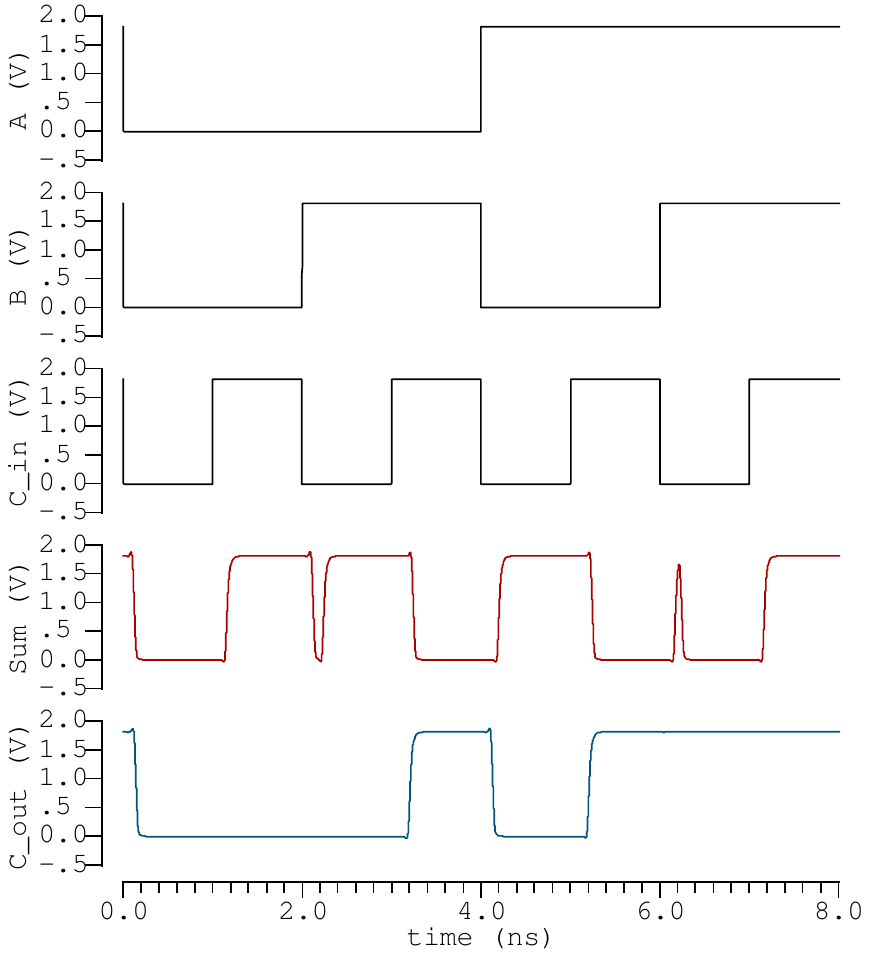}
\caption{Transient response of full adder circuit for all possible combinations. Half adder is designed using \XOR gate for SUM and \AND for CARRY out.}
\label{fulladdertran}
\end{figure}

The performance is analysed for full adder circuit as shown in Fig.~\ref{fulladdertran}. Transient analysis is simulated for all the possible combinations of the circuit. The Top three waveform shows signal A, B and Cin and the result is plotted or Sum and Carry Signal as shown in Fig.~\ref{fulladdertran}

The major problem with the linear ion drift model is the non-stabalized output parameters. That is the level degradation The level degradation can be approximated due to the voltage divider circuit and thus the TEAM model is preferred to raise the output level of the signal.

In full adder transient analysis there are very slight glitches that are due to CMOS technology used. Memristors alone provides near ideal transient response for the circuit. 

Delay and Rise time/fall time of the circuits are extracted for the given transient response is Tr = 43.71 ps, Tf = 22.43 ps. The delay calculated is 98 ps for half adder circuit. The normalised power for the half adder for all possible combinations is 8.07. For full adder circuit, the Tr = 82.12 ps and Tf = 34.1 ps. The delay analysed is 213.3 ps for full adder circuit. The normalised power for the full adder for all the possible combinations is 17.87 $\mu$W.

The four bit adder is designed using 4 one bit full adders and then two four bit adders are cascaded to implement the 8 bit adder. The performance parameters for 8 bit Full adder are extracted as Tr = 114.2 ps, Tf = 78.7 ps, Delay = 371.3 ps for worst case and normalised power = 52.7 $\mu$W.

\begin{figure}[!t]
\centering
\includegraphics[width=3.5in]{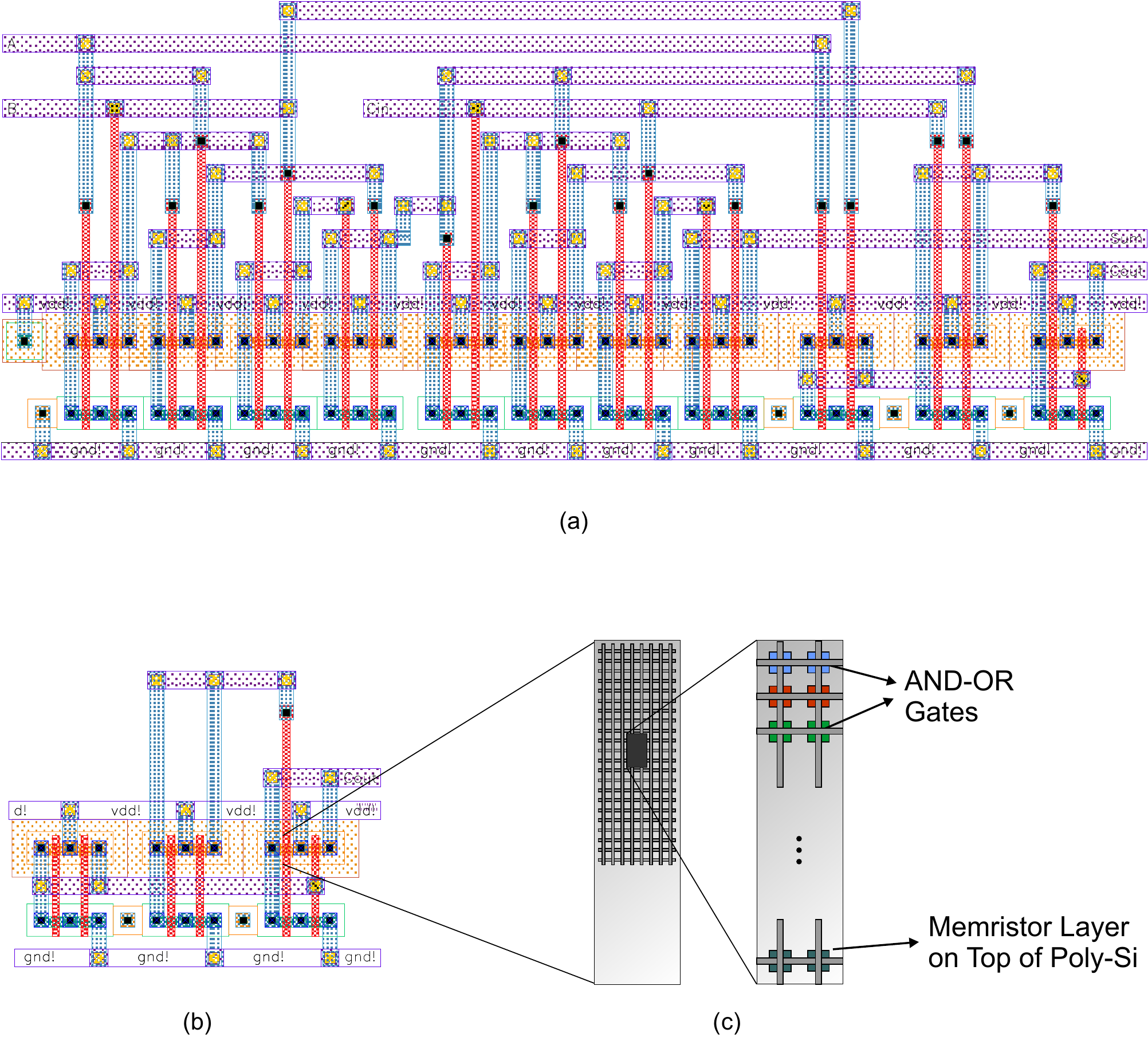}
\caption{(a). Layout of CMOS based full adder circuit using standard 36 transistor approach. Due to the design complexity inputs A, B and Cin, outputs SUM and Cout are connected with Metal2 layer. Power and ground rails are also connected with Metal2 layer. (b). Layout of CMOS layer if used for hybrid memristor-CMOS logic base implementation. The area of CMOS layer can be reduced by approximately 4x. (c). Possible layout design of memristor layer on top of polysilicon layer of MOSFET. The connections can be made using vias. Many \AND and \OR gates can be implemented just on the polysilicon layer of a MOSFET.}
\label{adder_memristor_layer}
\end{figure}

For Hybrid Memristor-CMOS logic, the layout for the circuits designed completely with CMOS logic and for the possible implementation of Hybrid Memristor-CMOS logic. Fig.~\ref{adder_memristor_layer} shows the implementation using CMOS 180 nm process technology. For the sake of comparison, the \AND gate required 6 MOSFETs but in the case of hybrid memristor-CMOS logic, there is no any MOSFET required. Hence, even for n-input \AND or \OR gate, there is much saving in area as memristors can be implemented over the top of MOSFETs and through interconnects the gates can be connected together. \index{Hybrid Memristor--CMOS Logic!Layout Design}

The layout of CMOS full adder is shown in Fig.~\ref{adder_memristor_layer}. In the Fig.~\ref{adder_memristor_layer}(a), the full adder circuit is shown. In Fig.~\ref{adder_memristor_layer}(a), the reduced number of MOSFETs are only required for the implementation of full adder circuit. Fig.~\ref{adder_memristor_layer}(c) shows the possible layout of hybrid memristor-CMOS logic circuit with memristor layer on top of CMOS layer. 

Memristors are connected on the poly-silicon layer on top of MOSFET's gate. The connection with the gate can be done with the help of vias. Two memristors are required for the computation of \AND and \OR function, that is shown in Fig.~\ref{adder_memristor_layer}(c), lot of complex functions can be implemented that take the area similar to one MOSFET. Hence, it can further help in the reduction of area and implement more logic functions per unit area of chip.

\begin{figure}[!t]
\centering
\includegraphics[width=3.5in]{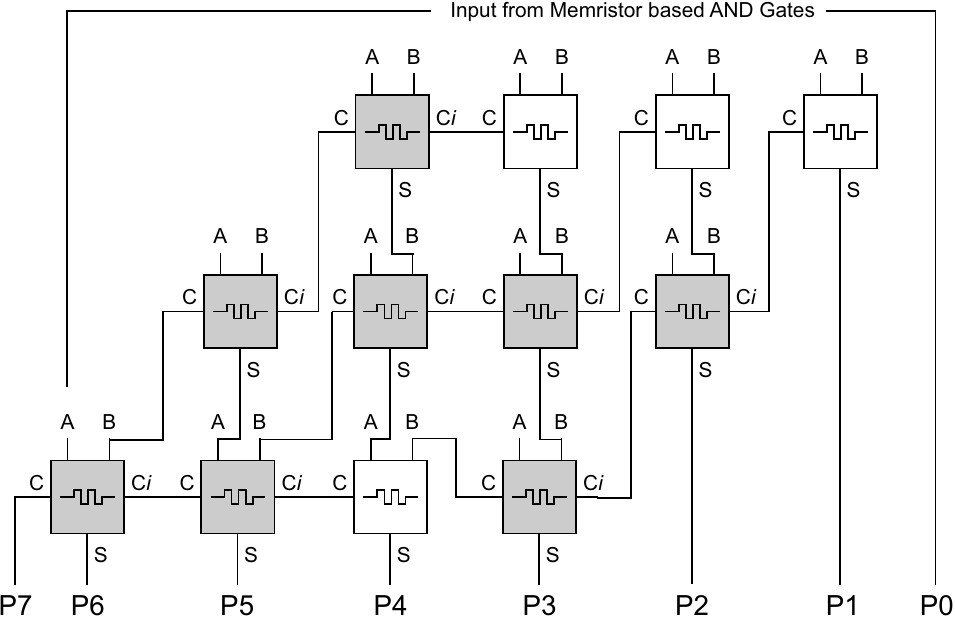}
\caption{Proposed Multiplier architecture using hybrid memristor-CMOS logic as described. All the adders and gates are defined using MeMOS logic.}
\label{multiplier}
\end{figure}

\begin{table}[!b]
\centering
\caption{Performance analysis of logic gates and adder circuits using Hybrid Memristor-CMOS logic.}
\label{hybrid_parameters}
  \begin{tabular}{@{\ \ } lcccr @{\ \ }}
    \toprule
  	Logic	  & Rise Time & Fall Time & Delay & Dyn+Sta\\
    Component & (T\textit{r}) 		  & (T\textit{f})  & \textit{d}	 & Power \\ 
    \midrule
    NOT 		& 23.3 ps & 14.1 ps & 18.70 ps & 0.5 $\mu$ \\ 
    AND 		& 02.2 ps & 00.8 ps & 1.50 ps & 1.50 $\mu$\\ 
    OR 			& 02.1 ps & 00.8 ps & 1.45 ps & 1.51 $\mu$\\ 
    NAND 		& 23.4 ps & 19.1 ps & 21.25 ps & 1.82 $\mu$\\ 
    NOR 		& 28.1 ps & 14.2 ps & 21.15 ps & 1.83 $\mu$\\ 
    XOR 		& 40.4 ps & 20.8 ps & 30.60 ps & 2.08 $\mu$\\ 
    XNOR 		& 40.1 ps & 22.1 ps & 31.11 ps & 2.41 $\mu$\\ 
    Half Adder 	& 43.7 ps & 22.4 ps & 98.05 ps & 8.07 $\mu$\\ 
    Full Adder 	& 82.1 ps & 34.1 ps & 212.3 ps & 17.87 $\mu$\\ 
    8bit FA 	& 114.2 ps & 78.7 ps & 371.3 ps & 52.71 $\mu$\\ 
    \bottomrule
  \end{tabular}
\end{table}

\begin{table}[!b]
\centering
\caption{Performance analysis of logic gates and adder circuits using CMOS logic.}
\label{cmos_parameters}
  \begin{tabular}{@{\ \ } lcccr @{\ \ }}
    \toprule
  	Logic	  & Rise Time & Fall Time & Delay & Norm.\\
    Component & (T\textit{r}) 		  & (T\textit{f})  & \textit{d}	 & Power \\ 
    \midrule
    NOT 		& 23.3 ps & 14.1 ps & 18.70 ps & 5.41 $\mu$ \\ 
    AND 		& 35.0 ps & 18.7 ps & 26.85 ps & 19.28 $\mu$ \\ 
    OR 			& 29.4 ps & 27.6 ps & 28.51 ps & 19.63 $\mu$ \\ 
    NAND 		& 47.8 ps & 20.9 ps & 34.35 ps & 10.69 $\mu$ \\ 
    NOR 		& 50.7 ps & 17.2 ps & 33.90 ps & 10.88 $\mu$ \\ 
    XOR 		& 83.9 ps & 48.3 ps & 66.02 ps & 47.81 $\mu$ \\ 
    XNOR 		& 78.8 ps & 50.4 ps & 64.61 ps & 43.61 $\mu$ \\ 
    Half Adder 	& 85.2 ps & 47.8 ps & 126.2 ps & 58.32 $\mu$ \\ 
    Full Adder 	& 96.4 ps & 54.2 ps & 342.7 ps & 117.3 $\mu$ \\ 
    8bit FA 	& 183.1 ps & 106.5 ps & 586.2 ps & 0.98 m \\ 
    \bottomrule
  \end{tabular}
\end{table}

\begin{table}[!t]
\centering
\caption{MOSFET and memristor count for logic gates using hybrid memristor-CMOS logic and CMOS logic.}
\label{mos_count}
  \begin{tabular}{@{\ \ } lcccccccc @{\ \ }}
    \toprule
	Device & \NOT & \AND & \OR & \NAND & \NOR & \XOR & \XNOR & \texttt{BUF} \\ 
    \midrule
    \multicolumn{9}{c}{Hybrid Memristor-CMOS Logic}\\
    MOSFETs  & $2$ & $0$ & 0 & 2 & 2 & 4 & 4 & 4 \\
    Memristors & 0 & 2 & 2 & 2 & 2 & 6 & 6 & 0 \\
    \midrule
    \multicolumn{9}{c}{CMOS Logic}\\
    MOSFETs  & 2 & 6 & 6 & 4 & 4 & 12 & 12 & 4 \\
    \bottomrule
  \end{tabular}
\end{table}

\begin{table}[!b]
\centering
\caption{MOSFET and memristor count for adder circuits using hybrid memristor-CMOS logic and CMOS logic.}
\label{mos_count2}
  \begin{tabular}{@{\ \ } lccc @{\ \ }}
    \toprule
	Device & Half Adder & Full Adder & 8-bit Adder \\ 
    \midrule
    \multicolumn{4}{c}{Hybrid Memristor-CMOS Logic}\\
    MOSFETs	   & 8 & 16 & 128 \\
    Memristors & 8 & 18 & 144 \\
    \multicolumn{4}{c}{CMOS to Memristor Layer Transitions}\\
    VIAs    & 5 & 10 & 80 \\
    \midrule
    \multicolumn{4}{c}{CMOS Logic}\\
    MOSFETs    & 14 & 34 & 272 \\
    \bottomrule
  \end{tabular}
\end{table}

\begin{table}[!t]
\centering
\caption{Comparison of parameters of hybrid memristor-CMOS logic with CMOS Technology}
\label{table_comparison}
  \begin{tabular}{@{\ \ } lccc @{\ \ }}
    \toprule
	Parameter & Half Adder & Full Adder & 8-bit Adder \\
    \midrule
	Area Utilisation & 57.2\% & 47\% & 47\% \\
	Included BUFFERS & \multicolumn{3}{c}{50\% of total area} \\
	Performance	     & \multicolumn{3}{c}{2.2x CMOS}\\
	Complexity		 & \multicolumn{3}{c}{Greater than CMOS}\\
    \bottomrule
  \end{tabular}
\end{table}

Hybrid Memristor-CMOS logic based gates and adder architectures are designed using Cadence Virtuoso and the performance parameters are analysed. The results are acquired for all logic gates, half adder, full adder and 8-bit full adder. The comparison is done with current generation CMOS technology. The parameters of MOSFETs, Memristors and supply voltages are kept constant for fair comparison in between these two logic families. The advantages and disadvantages of these logic families are discussed in previous sections. The IMPLY logic family is not included in analysis due to the highest number of computational steps required for any boolean functions.

The extracted performance parameters of various gates and adder circuits are given in Table~\ref{hybrid_parameters} for Hybrid memristor-CMOS logic and for CMOS logic the parameters are given in Table~\ref{cmos_parameters}. The required number of transistors in both logic families are given in Table~\ref{mos_count} and \ref{mos_count2} for hybrid memristor-CMOS logic and for CMOS logic respectively. The parameter comparison of adder circuits with CMOS logic is shown in Table~\ref{table_comparison} 

\subsection{Comparison between logic families}\index{Comparison between Logic Families}
\subsubsection{Speed}
In general, the total calculation time for any logic computation is Hybrid memristor-CMOS logic $<$ CMOS logic.
From the parameters like rise time, fall time and delay, hybrid memristor-CMOS logic seems prominent than CMOS logic.
\NOT gate is used for \NOT function in both the logic families, the maximum speed limit in hybrid memristor-CMOS logic is due to \NOT gate.

\subsection{Area Utilisation}\index{Area Utilisation}\index{Comparison between Logic Families!Area Utilisation}
For the assumption that memristors are smaller than MOSFETs, thus the area utilisation is hybrid memristor-CMOS logic $<$ CMOS logic.
This new logic ushers in the area saving because the memristors considered for analysis is of width = 3 nm which is way smaller than width = 180 nm of a MOSFET.
Memristors can possibly be implemented on the polysilicon layer of a MOSFET, thus a single MOSFET can be a house for many memristors. Many complex functions can be implemented under the area of a single MOSFET whereas in CMOS logic much more transistors are required for the computation of similar function.
The layout of full adder is demonstrated to compare the number of MOSFETs required for both logic families. There is a reduction of around 47\% in area if Hybrid Memristor-CMOS logic is used.

\subsection{Controller Complexity}\index{Controller Complexity}\index{Comparison between Logic Families!Controller Complexity}
The only logic family that required an external read/write controller is the IMPLY logic family. Because, IMPLY logic family requires internal states for read/write operation, there is a need to convert this internal states viz. resistances into its equivalent logic states to implement with CMOS. Hence, controllers are required that takes a lot of area and can further reduce the speed of IMPLY logic. 

For Hybrid memristor-CMOS logic and CMOS logic there is no any extra read/write circuit is required. The primary purpose of this hybrid memristor-CMOS logic is to build a new logic family that can be integrated with current generation of CMOS technology.

\subsection{Power Dissipation}\index{Power Dissipation}\index{Comparison between Logic Families!Power Dissipation}
Generally, current generation CMOS process consumes more power due to the constant connection of supply voltage that induce static leakage. However, in hybrid memristor-CMOS logic, static leakage is there. The case is not same with IMPLY logic, however, controllers will be having MOSFETs that means if seen from a broader viewpoint, complete architecture including read/write circuits, there is a static power dissipation~\cite{chan1987impact}.

Static power dissipation can be further reduced by well known techniques like clock gating or multi threshold CMOS. Static power dissipation is not a major concern for the average power, the maximum power dissipation is the dynamic power, that is due to the transitions. Power is computed for signal that have maximum transitions. The power is normalised for memristors because the computation of power is not the same as of CMOS~\cite{manem2012stochastic,shin2011reconfigurable}.

\subsection{Versatility}\index{Verstility}\index{Comparison between Logic Families!Versatility}
Till now CMOS logic is the most versatile logic, as it offer many different ways to design the circuit or architecture according to the needs of complete system. But as with the Hybrid memristor-CMOS logic, the versatility increases, because the memristor layer on top of CMOS layer, adds advantage to compute some of the functions of a separate layer and the results can be taken directly from that layer.

In the previous section, the full adder is designed in the same way, the SUM is computed with the help of CMOS and memristor layer, through VIAs the signal gets exchanged but the CARRY is computed entirely on memristor layer, because the \AND and \OR gates are required for CARRY can be implemented in memristors only.

\subsection{Further Improvements}\index{Further Improvements}\index{Comparison between Logic Families!Further Improvements} \index{Hybrid Memristor--CMOS Logic!Further Improvements}
In Hybrid memristor-CMOS logic, the speed is mostly dependent on the working voltages provided that are mostly determined by the threshold voltages. Hence, memristors with higher current threshold value are faster but consumes more power.

In terms of the area utilisation, if the speed is a primary concern, current threshold should be high, thus there is a requirement of more BUFFERs that affects the area requirement of hybrid memristor-CMOS logic.  

Power is usually dominated by the applied voltages, that can be chosen as per the current threshold. One method to reduce power dissipation is to add more BUFFERs with CMOS logic gates, BUFFERs can reduce static power dissipation as well as dynamic power dissipation as BUFFERs eliminated the glitches. But the compromise in terms of area is there. Static power can be completely eliminated by adding BUFFERs after each successive hybrid memristor-CMOS logic state.

\section{Conclusion}
In this work, we have studied the models of memristors and its application in logic circuits. We used TeAM model to implement Hybrid Memristor-CMOS (MeMOS) based logic architectures. The degradation factor using linear ion drift model is also considered and thus by using TeAM model. Logic gates are designed with CMOS 180 nm process technology. Adder circuits are designed and the performance is analysed and compared with current generation CMOS 180 nm technology. Area utilisation using IMPLY logic and proposed logic is also compared. Possible layout configuration of MeMOS logic is also described. This paper opens the possibility of newly developed memristor for logic circuits. Based on the excellent performance of adder circuits, this work can be extended further on complex logic architectures like multipliers and many more.
\bibliographystyle{IEEEtran}
\bibliography{thesismemref}

\begin{IEEEbiography}[{\includegraphics[width=1in,height=1.25in,clip,keepaspectratio]{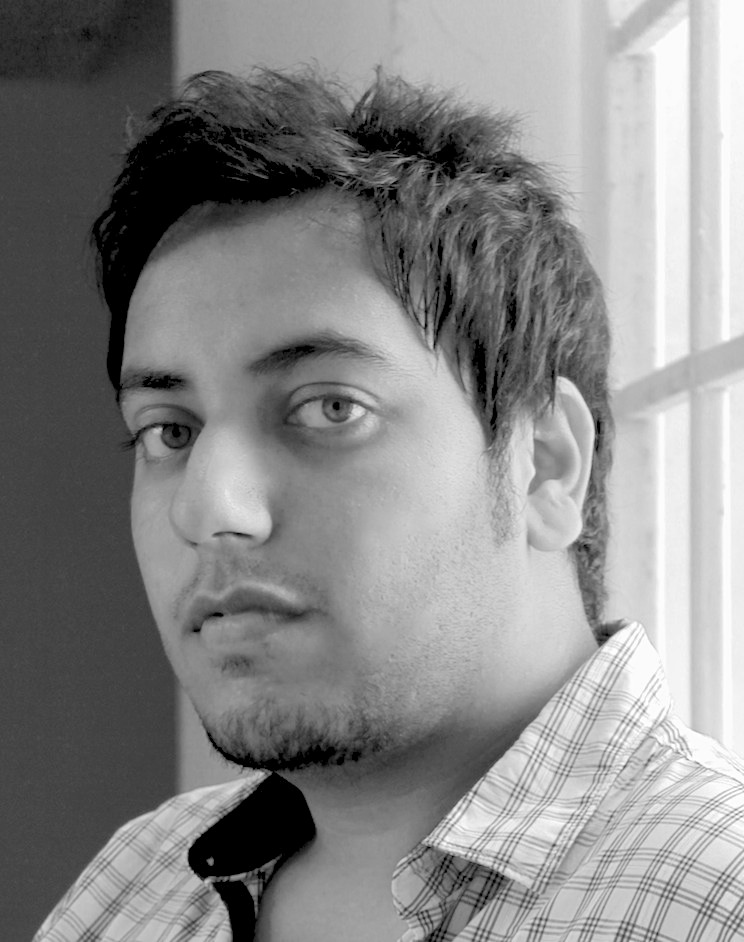}}]{Tejinder Singh}
(S'12, M'14) received his 3-year Technical Diploma (2007) in electronics and communication engineering from Ramgarhia Polytechnic, PB, India. B.Tech.~(2010) degree in electronics and communication engineering and M.Tech.~(2014) degree in VLSI from Lovely Professional University, PB, India. From 2010 to 2012 he was an electrical and computer engineer in an industrial sector. Currently, he is an assistant professor in discipline of electrical and electronics, Lovely Professional University from 2014. His research interests include the designing, modeling and characterization of RF MEMS devices in microwave and millimeter-wave circuits, CMOS-MEMS integrated circuits and systems, VLSI architectures, Memristor/Memristive systems. He has authored or coauthored over 30 publications in peer-reviewed journals and international conferences. He has served as the technical program committee board member of more than 34 international conferences/symposiums/workshops. 

Mr. Singh is listed in the 33\textsuperscript{rd} edition of Marquis Who's Who in the World (2015). He is the recipient of many awards including the 2013 IEEE Asia-Pacific region PG research paper award from IEEE, USA, 2013 IEEE M. V. Chauhan award, IEEE B.R.V. Vardhan award from IEEE, India. He has received best paper awards at the 4\textsuperscript{th} International Conference in Computing, Communication and Networking Technologies (ICCCNT 2013), TN, India, at the International Conference of Control, Computing, Communication and Materials (ICCCCM 2013), UP, India and at International Conference on Computing Sciences, PB, India (ICCS 2013). His research has won numerous accolades from Lovely Professional University at the School and University level. He is a member of IEEE, IEEE MTT-S, IEEE COMSOC, IEEE EDS, ACM, ASME, SPIE, IOP Science, NSPE, Engg. AUS., Engg. IRE., ISOC and many other international professional bodies.
\end{IEEEbiography}
\end{document}